\documentclass[11pt]{elsarticle}

\pdfoutput=1

\makeatletter
\def\ps@pprintTitle{%
  \let\@oddhead\@empty
  \let\@evenhead\@empty
  \let\@oddfoot\@empty
  \let\@evenfoot\@oddfoot
}
\makeatother

\usepackage{url}
\usepackage{breakurl}
\usepackage[breaklinks,
            colorlinks = true,
            linkcolor = blue,
            urlcolor  = blue,
            citecolor = blue,
            anchorcolor = blue]{hyperref}

\usepackage{lineno}

\usepackage{graphicx}
\usepackage[margin=1.25in]{geometry}
\usepackage[usenames,dvipsnames]{color}

\newcommand\snowmass{\begin{center}
\rule[-0.2in]{\hsize}{0.01in}\\\rule{\hsize}{0.01in}\\
\vskip 0.1in Submitted to the  Proceedings of the U.S. Community Study\\ 
on the Future of Particle Physics (Snowmass 2021)
    \vskip 0.05in
    {\it Snowmass 2021 CEF03 Diversity, Equity \& Inclusion}\\
\rule{\hsize}{0.01in}\\\rule[+0.2in]{\hsize}{0.01in} \\
\end{center}}

\usepackage[firstpage=true]{background}
\backgroundsetup{contents={\parbox{6.5in}{\snowmass}}, scale=1,placement=top,opacity=1,color=black,position={3.25in,1.2in}}

\usepackage{fancyhdr}
\fancypagestyle{plain}{%
  \fancyhf{}%
  \fancyhead[C]{}
  \fancyfoot[C]{\thepage}
}

\fancypagestyle{empty}{%
  \fancyhf{}%
  \fancyhead[C]{{\it Snowmass 2021 CEF03 Diversity, Equity \& Inclusion}}
  \fancyfoot[C]{\thepage}
}
\begin{document}

\begin{frontmatter}


\title{Why should the U.S. care about high energy physics in Africa and Latin America?}

\author[add1]{K\'et\'evi A. Assamagan\corref{cor1}}
\ead{ketevi@bnl.gov}
\author[add2]{Carla Bonifazi}
\author[add3]{Johan Sebastian Bonilla Castro}
\author[add4]{Claire David}
\author[add5]{Claudio Dib}
\author[add6]{Luc\' ilio Dos Santos Matias}
\author[add7]{Samuel Meehan}
\author[add8]{Gopolang Mohlabeng}
\author[add9]{Azwinndini Muronga}

\cortext[cor1]{Corresponding Author}

\address[add1]{Brookhaven National Laboratory, Physics Department, Upton, New York, USA}
\address[add2]{ICAS-ICIFI-UNSAM/CONICET, Argentina and Universidade Federal do Rio de Janeiro, Brazil}
\address[add3]{University of California, Davis, USA}
\address[add4]{York University, Canada \& Fermi National Accelerator Facility, USA}
\address[add5]{Universidad T\'ecnica Federico Santa Maria
Valparaiso, Chile}
\address[add6]{Division of Medical Physics, Department of Physics, Eduardo Mondlane University, Maputo, Mozambique}
\address[add7]{CERN, Switzerland, 2021 -- 2022 AAAS Science \& Technology Policy Fellow, USA}
\address[add8]{Department of Physics and Astronomy, University of California, Irvine USA}
\address[add9]{Faculty of Science, Nelson Mandela University, Gqeberha, South Africa}

\begin{abstract}
\noindent Research, education and training in high energy physics (HEP) often draw international collaborations even when priorities and long term visions are defined regionally or nationally. Yet in many developing regions, HEP activities are limited in both human capacity and expertise, as well as in resource mobilisation. In this paper, the benefits -- to the U.S. HEP program -- of engagements with developing countries are identified and studied through specific examples of Africa and Latin America; conversely, the impact of HEP education and research for developing countries are also pointed out. In the context of the U.S. strategic planning for high energy physics, the authors list recommendations on investments that will benefit both developed and developing nations.
\end{abstract}

\begin{keyword}
HEP \sep Snowmass \sep Community Engagement Frontier \sep CEF03 \sep Diversity, Equity and Inclusion \sep developing countries
\end{keyword}

\end{frontmatter}


%


\newpage

\section{Why should the U.S. care about physics in developing countries?}
\label{sec:why}
\noindent Throughout the last century and a half, the U.S. has invested human, financial, and social capital throughout the world. Benefits to the U.S. can be seen through many perspectives, one such approach is to consider the benefits to the U.S. as affecting its domestic or foreign relationships. Strengthening domestic relationships means developing U.S. institutions, this includes growth of capabilities of U.S. National Laboratory programs, the success of public and private universities, versatility of the workforce, etc. Regarding international endeavors, the U.S. benefits from healthy relationships between foreign countries and their institutes with flow of information, technology, and many other resources. In order to have a steady flow, it is important to maintain a positive relationship with the world. We argue that the U.S. can benefit from supporting the progress of fundamental science in developing counties in all of these facets of domestic and foreign relationships.

The efforts of the Peace Corps or the U.S. Agency for International Development (USAID) fall within these investments~\cite{PeaceCorps, USAID}. The ultimate objective is not only to show the best of American values (or furthering American interests), but also to help reduce poverty and make progress well beyond assistance~\cite{USAID}. In the context of the USAID program, the African-American Institute (AAI)~\cite{aai} has played the role of engaging with African countries to administer the U.S. aid. In addition, the AAI promotes development in Africa through a wide range of activities and programs~\cite{aai}. The Fulbright program~\cite{fulbright} is another example of a U.S.-supported international exchange framework to promote inter-cultural understanding and skill development in developing countries. These programs strongly encourage international beneficiaries who go to the U.S. for training and high education to return to their home countries to help; this is often enforced through the "two-year home-country residency requirement" that the recipient of such support must satisfy at the completion of their stay in the U.S.~\cite{USAID, fulbright}. Other international aids are provided through foundations of non-governmental agencies such as the Bill and Melinda Gates Foundation~\cite{BillMelinda}, etc.

Many education and research institutes in the U.S. offer international studies or programs, not just for students, but as well for professionals and businesses to improve international engagements and cultural awareness and sensitivities~\cite{yale, ucla}---these programs are funded in parts through a Title VI Federal Area Studies grant sponsored by the U.S. Department of Education. The Title VI is a provision of the 1965 Higher Education Act, "to support area studies that serve as vital national resources for world regional knowledge and foreign languages. This program reflects the special priority placed by the federal government on foreign language and area studies, especially with respect to diplomacy, national security, and trade competitiveness"~\cite{title6}.

The United Nations (UN) General Assembly has proclaimed the year 2022 as the "international year of basic sciences for sustainable development, to improve the quality of life for people all over the world"~\cite{UN}. The UN proclamation resulted from efforts made by the international scientific communities, including CERN~\cite{CERN}, IUPAP~\cite{IUPAP} and others under the auspices of UNESCO~\cite{UN, UNESCO}; this serves to support constructive engagements with developing countries, to improve their physics education and research programs, for the benefit of all humankind.

The foregoing case study suggests that the U.S. should care about physics in developing countries to support national interests, values and ideals, with the collateral benefit of seeding self-sustaining development. 

The rest of this article is structured as follows; in Section~\ref{sec:current}, we discuss the current known status of HEP in some developing nations. Section~\ref{sec:invest} is left for discussion on how investment in science research in the developing world may be improved. In Section~\ref{sec:outcome}, we discuss the expected outcomes for the U.S. and developing nations and finally we conclude and make some recommendations in Section~\ref{sec:conc}.

\section{What is the current status of HEP in Africa and Latin America?}
\label{sec:current}

\subsection{Involvements of Africa and Latin America in large international collaborations}

Large scale activities in HEP draw on global efforts, where groundbreaking discoveries can be achieved through international collaborations. However, many countries have not contributed equitably to international endeavors in HEP, e.g. countries classified as developing nations. Various indices are used for such classifications, for example the Gross National Income \emph{per capita}, the Human Asset Index or the Economic Vulnerability Index, as defined by the United Nations or the World Bank~\cite{UN-indices, WorldBank-indices}. Other metrics to assess developing nations include the Human Development Index, the Multidimensional Poverty Index, and the Happy Planet Index~\cite{Alternate-indices}. Ultimately, activities in HEP rely on substantial base of development in economics~\cite{EconoSectors}.
\begin{figure}[!htpb]
\begin{center}
\includegraphics[width=\textwidth]{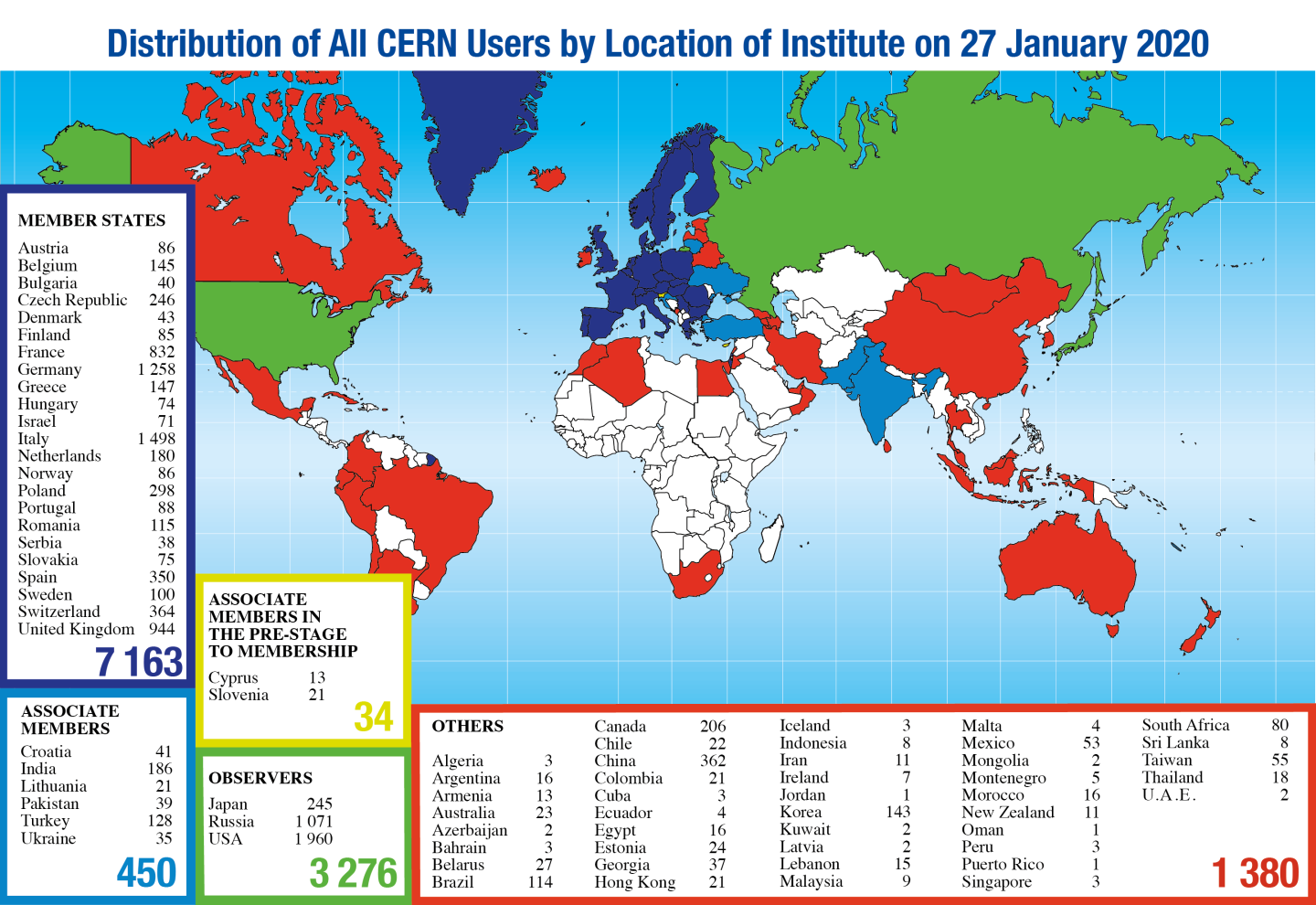}
\end{center}
\caption{Distribution of CERN~\cite{CERN} users by location of their institutes. Users are not necessarily originating from the country of the university or laboratory they are affiliated with. Less than 5\% of CERN users are associated with a developing nation.}
\label{fig:cern-users}
\end{figure}
Figure~\ref{fig:cern-users} shows the countries with institutes working for the European Organization of Nuclear Research (CERN)~\cite{CERN, cern-map} along with the number of affiliated scientists in each country, denoted "users". Users do not necessarily originate from the institute's country of their affiliation. In total, less than 5\% of users are associated with a developing nation.~\footnote{If we include China, the number goes to 8\%.}  The geographic diversity of CERN users is broadly applicable to the HEP community or other centers of major HEP engagement such as Fermilab. 
In Africa only four countries -- all developing nations -- have institutes working for CERN, with a share of users less than 1\%. Although Figure~\ref{fig:cern-users} suggests that the situation is relatively improved for South American institutes; their relative CERN-user engagement is below 2\%. In addition, according to the American Physical Society (APS) member survey, African and Latin American memberships are only 1\% and 3\% respectively \cite{APS-taskforce}.
\begin{figure}[!htpb]
\begin{center}
\includegraphics[width=\textwidth]{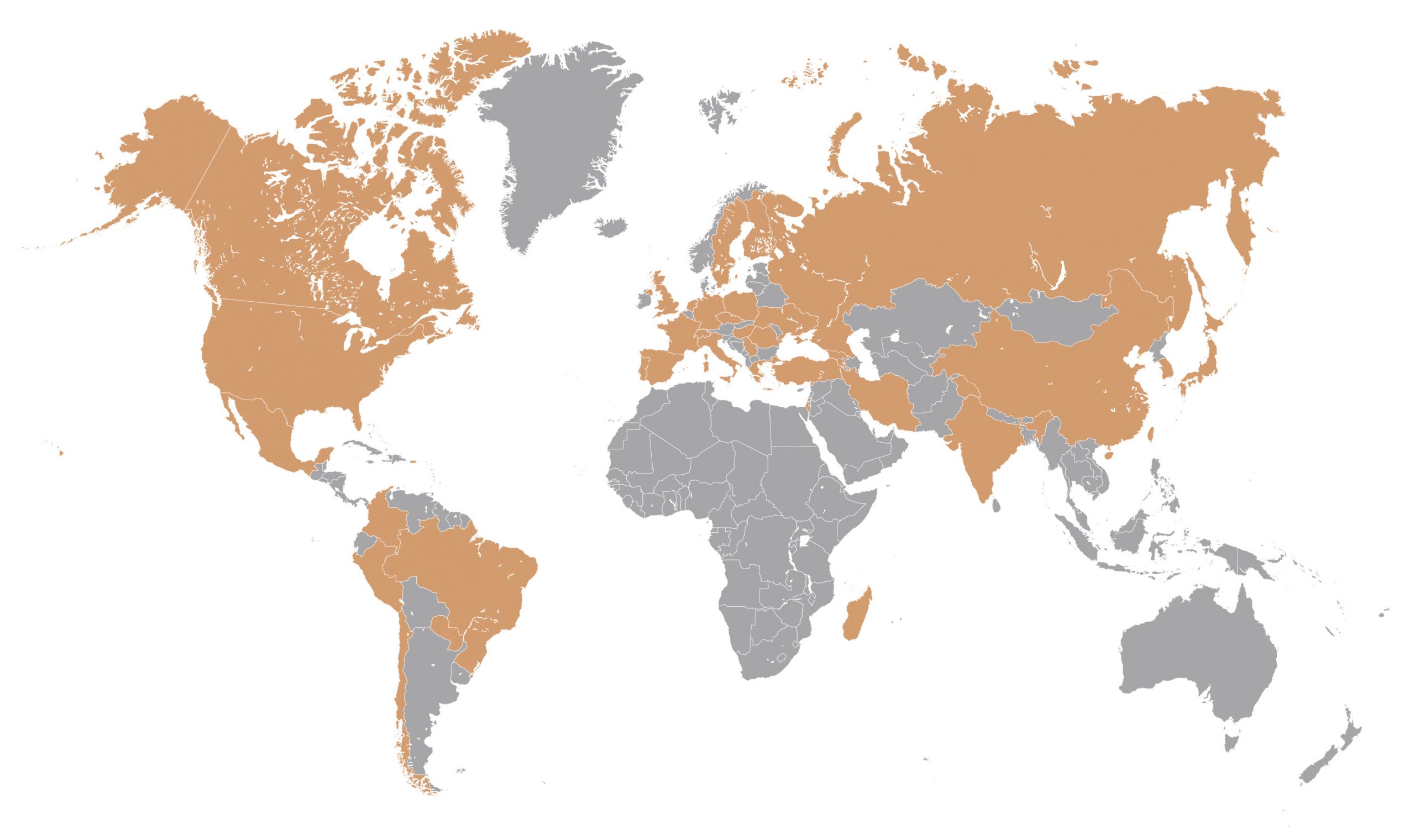}
\end{center}
\caption{The Deep Underground Neutrino (DUNE) collaboration by country participation. Madagascar is the only African country in this collaboration as of 2022.}
\label{fig:dune}
\end{figure}

As shown in Figure~\ref{fig:dune}, Madagascar is the only African country involved in the Deep Underground Neutrino Experiment (DUNE); the Malagasy collaboration is 0.3\% of the size of DUNE~\cite{dune}.

The focus of the analysis presented in this paper is limited to countries in Africa and Latin America. However, the findings may be generalized to other developing nations around the world. 

\subsection{The case of Africa}
In Africa, HEP activities are centered in the North and South; however, there are some random positive signals from the vast array of countries on the continent~\cite{Allotey1996}. Some scientists from Africa and Latin America are affiliated to research institutes abroad and might eventually return to their home countries; until they do, their institutional contributions to HEP are not reflecting in countries of origin. 

In North Africa, countries such as Morocco and Egypt have the largest presence. Experimentally, the two countries have invested in the ATLAS and CMS collaborations respectively. Morocco joined ATLAS in 1996 and, at the time of writing, has 52 active members. Egypt joined CMS as a full member, and ALICE as associate member, in 2012. Currently, Egypt-CMS has 32 members. The expertise in both countries range from physics analyses to computing and detector performance, operation and upgrade. On the theory side, in both countries there are about 20 members (including Ph.D. and MSc. students) who focus on the Standard Model and Beyond the Standard Model physics,  quantum gravity, string theory, astrophysics and cosmology~\cite{HEP-Africa}. Algeria is part of the LHCb experiment---with four active members from the Laboratory of Mathematical and Subatomic Physics,  Constantine---as an associate member to Universidade Federal do Rio de Janeiro (UFRJ), Rio de Janeiro, Brazil~\cite{LHCb}. Morocco has been involved in ANTARES~\cite{Antares} and the development of next generation of marine telescopes known as KM3Net~\cite{KM3Net}. 

In the Southern region of the continent, HEP is dominated by countries such as South Africa and Madagascar. South Africa is part of the ATLAS collaboration, with a total of 40 active members, and has been a member of ALICE (since 2001) and ISOLDE collaborations, with 15 and 21 active members respectively. 
In  2008, South Africa launched the SA-CERN Consortium \cite{HEP-Africa}, which provides funding for students and researchers to travel to CERN for research and collaboration. In addition to Experimental programs, SA-CERN has an SA-CERN/Theory division which caters for theoretical physics in CERN related physics. South Africa is also partnering with the Joint Institute for Nuclear Research (JINR) in Russia. The University of Antananarivo in Madagascar has recently joined the DUNE collaboration, with 9 active members involved in physics analysis and detector development. HEP theory on the southern part of the continent is dominated by South Africa, which has formed professional bodies such as the SA-CERN theory consortium \cite{HEP-Africa}, the SA-JINR theory collaboration and the National Institute of Theoretical and Computational Sciences (NITheCS)~\cite{SA-JINR}. Research expertise in these institutes ranges from High Energy Nuclear theory to HEP Phenomenology and more~\cite{HEP-Africa, SA-JINR}. The reconfiguration of NITheP to NITheCS means that there are great opportunities for trans-disciplinary research, in particular HEP and astroparticle physics. Through an initiative with the IAEA~\cite{iaea}, South Africa is serving as a hub to the whole Africa for the "small" facilities in nuclear and particle physics (accelerators); and many African countries are involved, not only from the North and South. It is hoped that these initiatives will attract U.S. HEP and astroparticle community to partner with Africa.

The square kilometer array (SKA), mainly housed in Southern Africa is serving as a catalysis for the collaboration between African astroparticle physicists and HEP communities on the continent. There could be leveraging of access to facilities between African particle physicists and U.S. astrophysicists and visa versa.

Despite low participation of African in HEP, there are notable scholars from Africa---or of the African diaspora---who have contributed in international HEP, especially in the theory sector. 

The African-American Institute organizes an annual international conference about the state of education in Africa~\cite{AAI-SOE}. The conference brings together stakeholders and practitioners worldwide to discuss mechanisms for strengthening high education in Africa. The conference proceedings not only summarizes the discussions, but also shares knowledge and recommendations with the broader audience; this conference offers a yearly snapshot of the state of high education in Africa.

Some physics programs, supported partially through international engagements includes the African School of Physics~\cite{ASP}, the African School for Electronic Structure Methods and Applications~\cite{ASESMA}, etc. Other physics programs are developed or coordinated through the ICTP~\cite{ICTP} or AIMS~\cite{AIMS}, and extensively detailed in this Snowmass~2021 contributed paper~\cite{Cecire2021}.

The activity reports of the African School of Physics suggest that only 15-20\% of the student participants have major concentrations in nuclear or high energy physics, and these HEP majors come mainly from North Africa and Southern Africa~\cite{ASP-reports}. These findings are supported by the institutional distributions of CERN users as shown in Figure~\ref{fig:cern-users}, and suggest that HEP education and research reflect national strategies in Africa and Latin America---the U.S. HEP community could influence these national priorities to better achieve the U.S. strategic agenda as defined by Title VI and increase the reach of HEP worldwide.

An educated African continent is a better Africa for everyone, U.S. institutes and efforts included. Africa's youth stands at about 60\% of the continent's population and is growing. Africa, being a youthful continent should have access to better education for stability and for poverty eradication, ending inequalities, and creating jobs for the youth. HEP plays an important role in the U.S. education. Access to HEP education by African youth will in turn assist in addressing the education challenges on the continent, not only will the education in the continent improve but also technology transfer will be accelerated. Ultimately, the brain-drain of Africa's talents will stop.

\subsection{The case of Latin America}
Latin America is constituted by countries with a wide range of economic and developmental capacities. Some countries, such as Argentina, Brazil, Chile, Colombia, and Mexico, have the population and public/private capital to fund research in large-ticket endeavors such as the LHC experiments. However, many Latin American countries do not have the history or in-country experience to lobby for larger expenses.

A common situation  for Latin American HEP researchers is to go abroad to earn an advanced degree, and to stay abroad due to lack of opportunities in their home countries. Theory research in this respect is much more transferable and feasible for smaller countries to support at their institutes. Hence, faculty in Latin American institutes tend to be skewed towards theory. Albeit, several Latin American countries experience brain-drain of experimentalists.
Without in-country expertise, it is difficult to create enough momentum to provide the space for HEP to flourish. To the U.S., this results in a more difficult task when investing in the development of these countries. To address this issue we, as a global community, need to identify and remove the obstacles to those countries' progress and inclusion in global collaborations. 

Here one must also consider that countries in Latin America not only have different levels of development, but also different policies and organizational structures for the financing of scientific research. Many countries in Latin America---Argentina, Bolivia, Brazil, Chile, Colombia Ecuador, Mexico, Paraguay, Peru and Venezuela---have HEP theory and experiment programs in ATLAS, CMS, ALICE, LHCb, RHIC, Belle II, High Luminosity LHC and future colliders. Latin America also has rich programs in neutrino physics and astroparticle physics and cosmology, with participation in many neutrino, dark matter, astroparticle and cosmology experiments, with a community of about a thousand physicists~\cite{HEP-LatinAmerica}.

Concerning large experimental HEP facilities, there are tandem accelerators for nuclear research in Argentina~\cite{tandar, tandar2}
and Brazil~\cite{Oliveira:2019yyx}, and a large Synchrotron Radiation lab in Brazil~\cite{LNLS}. However most major HEP facilities are in the field of astroparticle physics, established by world wide collaborations, such as the Pierre Auger Observatory in Argentina~\cite{AugerObs}, HAWC in Mexico~\cite{hawc}, 
CTA currently in construction in Chile~\cite{ctao}, SWGO in the process of selecting its site in South America~\cite{swgo}, and recently CONDOR~\cite{condor}, a new proposal of a high altitude observatory for cosmic and gamma rays, by a collaboration of scientists from UC Riverside in the U.S. and Chile. In addition, there is a proposal originated from the Latin American community to build ANDES, the first deep underground laboratory in South America, in a projected road tunnel to be built across the Andes mountains between Argentina and Chile~\cite{Bertou:2012fk, Dib:2015rma, andeslab2}.

The HEP community in Latin America has seen a steady growth in several countries, especially with the advent of the LHC~\cite{LASF4RI2021}. Several international support programs such as e.g. HELEN~\cite{HELEN}  have helped with funds for mobility and participation in experiments. In 2001 CERN established the first CERN Latin American School on High Energy Physics (CLASHEP)~\cite{CLASHEP}, a two-week event that takes place in a different Latin American country every other year, with students mainly -- but not exclusively -- from Latin America. Chile for example had no experimental HEP research before 2007, the year it joined the ATLAS Collaboration. Since then, several groups have been formed and participate in experiments both in the U.S. and Europe, with contributions that include fabricating technological components for detectors. Their Ph.D. programs have also grown in size and number, forming specialists that today work both within the country and abroad.

To better understand the mechanisms effective at developing HEP research capabilities in smaller Latin American countries, we can consider the maturing effort in Costa Rica to collaborate at CERN. Costa Rica is a relatively developed country in Central America of just over 5 million inhabitants. Arguably, its stable democracy and its commitment to education and peace makes it an ideal candidate for HEP research. Many unprecedented factors aligned in the past years that have allowed for a top-down creation of an experimental HEP group. Today however, only a handful of HEP-adjacent faculty have careers in the country who are relatively isolated from the modern HEP community.

Costa Rica became affiliated with CERN at the signing of an International Cooperation Agreement in 2014~\cite{CRCERN2014}, at the time in an attempt to join the ISOLDE collaboration. For various reasons, primarily due to lack of person-power and in-country expertise, the progress of the effort stalled. In 2017, the CERN-Costa Rica accord was ratified by the Costa Rican government~\cite{CRCERN2017}. In 2019, the interest in HEP research was rekindled with the visit of the President of Costa Rica, Carlos Alvarado Quesada, to CERN.  Since 2017, a group of three (then) Ph.D. students in physics, interested in the development of HEP in their home country, created relationships with the faculty behind the previous push and eventually organized a group academics that is now applying for membership to the LHCb collaboration. Efforts were made with the ATLAS and CMS collaborations but their large-ticket membership and detector operations financial commitments were too large to begin the dialogue with the Costa Rican funding agency. The flexibility of LHCb in financial contribution timeline, as well as its openness to in-kind contributions, is what made the difference for the burgeoning Costa Rican group.

In 2016, the Latin American HEP community began organizing a international forum to prioritize HEP efforts. In 2018 the efforts formalized with a Ministerial Declaration following the Third Ministerial Meeting of Science and Technology of Ibero-America explicitly recognizing a need for a prioritizing process, the Latin American Strategy for Research Infrastructures (LASF4RI). The objective of LASF4RI is to develop a roadmap to support Latin American collaborations and their impacts.~\cite{MinisterialDeclarationLASF4RI} Over the next two years the process continued gaining momentum and completed its first full cycle in 2020.~\cite{LA-Strategy}

In 2021 a \emph{Preparatory Group}, with delegates from ten Latin American countries as well as distinguished scientists representing other regions of the world, produced a LASF4RI-HECAP Strategy Document~\cite{LASF4RI2021}, which was submitted to a \emph{High Level Strategy Group} (HLSG). This document was written based on a detailed \emph{Physics Briefing Book} that resulted from community input (40 white papers submitted to the Preparatory Group), and a second HECAP topical workshop with broad regional representation that took place in July 2020. The Document concluded with 10 recommendations, four of which were highlighted by the HLSG: (i) participation from regional groups in the many current and planned projects located in Latin America should be increased, in order to help position them in prominent leadership roles, as well as benefit the younger generations of Latin American scientists; (ii) substantial incursion into underground physics experiments with the development of the ANDES facility should be a high priority for the region; (iii) the participation in international collaborations in both collider and neutrino physics experiments has been a key driver in the Latin American region, building-up experimental research groups in the region, which in many cases have taken on leading roles; these activities should be most strongly supported; (iv)
small and mid-scale regional projects of high scientific and technological impact can provide unique opportunities to the region to be innovative and assume leadership and should be supported with high priority.

\section{What has been the U.S. HEP community investments and how to improve?}
\label{sec:invest}

\noindent In the general context of international development, the HEP community has invested in activities in Africa and Latin America through field research, exchange programs, conferences and workshops, international schools and outreach~\cite{ICTP, AIMS, ASP, ASESMA, CLASHEP}, sabbaticals, and reviews of scientific agenda at the invitation of specific countries~\cite{SA-physics}. Specifically for the U.S. HEP community, many investments have been carried out through the structured programs mentioned in Section~\ref{sec:why} to support the activities described in Section~\ref{sec:current}.

Existing programs are often limited by financial resources and logistic at the venue---many of these programs seek financial support to maintain their services annually or biennially. Furthermore, venue capacity may also force the number of participants within manageable levels~\cite{AIMS, ICTP}. For example, the number of applicants at the African School of Physics exceeds 400 on average in each biennially edition and the selection rate is typically 25\%. However, financial considerations constrain the acceptance rate below 20\% before the venue capacity is reached; thus, 5 to 7\% of the selected students end up on waiting list and never get a chance to attend the school~\cite{ASP-reports, ASP2021-report}. COVID-19 has forced many activities into virtual platforms and this removes constraints of travel and venue capacity; however, online engagements require good internet connections, the lack of which affects the participation of folks in many developing regions---the virtual edition of the African School of Physics in 2021 illustrates the internet connectivity issues~\cite{ASP2021-report}.

The U.S. HEP community could increase its support to ensure that financial considerations do not constrain the rates of participation. In some cases, aid to improve the logistics at the venues or to boost internet connectivity may lead to increased levels of quality participation. In other cases, coverage for U.S.-based personnel, co-organizers, physicists and lecturers should come from different sources so that direct financial contributions to existing programs are used to maximize participation of Africans or Latin Americans.

Many large collaborations have high entrance fees and/or recurring long-term financial contributions that are agnostic to the entering institutes economic contexts (i.e. how much capital is possibly available to them). Collaborations like those at CERN expect the same contribution from an institute in a developed country as one in a burgeoning economy orders of magnitude smaller---flexibility in the formation of clusters of institutes, in the timeline and level of financial contributions, as well as openness in the types of contributions, would increase the reach of large scale HEP experiments in Africa and Latin America. Examples of such flexibility and openness that allowed developing countries to join large scale HEP collaborations include, but not limited to: Morocco has had traditional links to the European Union, in particular France. In the mid-1990s a small team from Casablanca and Rabat started collaboration with the ATLAS Grenoble group and this led to the full ATLAS membership of a Moroccan cluster in 1996~\cite{atlas}; Egypt signed a cooperation agreement with CERN in 2006 and joined CMS and ALICE in 2012~\cite{Egypt-CERN}; South Africa was associated to Brookhaven National Laboratory in 2008 before it became a member of ATLAS in 2010~\cite{SA-ATLAS}; in Section~\ref{sec:current}, we mentioned the case of Algeria associated to UFRJ Brazil, and that of Costa Rica, in LHCb; Madagascar became a member of the DUNE Collaboration in 2016~\cite{HEP-Africa}; recently, An-Najah National University became the first university in Palestine to join ATLAS in association with the University of Paris Saclay, France~\cite{An-Najah}.

\section{What are expected outcomes for the U.S. and developing countries?}
\label{sec:outcome}
In this paper, we've attempted to show that U.S. HEP engagements with Africa and Latin America fit within the U.S. national agenda. Defined through the Title VI Act, support for international development aims to protect U.S. interests and global competitive edge. This is the primary reason for the U.S. to care of physics, in particular HEP, in developing countries. Conversely, international development programs---particular in HEP---help foster basic education and research, and improve participation of developing countries in basic sciences for sustainable development towards a better quality of life for all people.

\section{Conclusions and recommendation}
\label{sec:conc}

Investment in physics in developing countries has the potential to benefit both developed and developing nations as summarized in Section~\ref{sec:outcome}. Considering the gains for U.S. national interests, it is recommended that:
\begin{itemize}
    \item The U.S. HEP community maintains the current engagements and increases investments in Africa and Latin America to improve the reach of HEP in these regions;
    \item U.S. universities, research labs, funding agencies help seed sustainable HEP activities in Africa and Latin America;
    \item U.S. universities and research labs encourage and support the participation of their personnel, faculties and research staffs in HEP education and research efforts of African and Latin American countries;
    \item U.S. institutes make regular financial contributions towards schools, workshops, conferences---HEP activities in general---in Africa and Latin America, to ensure a significant participation of students and faculties of these regions;
    \item U.S. institutes to partner with  Latin America and Africa in establishing bridge programs to facilitate the eventual re-integration of physicists from Africa and Latin America into their countries;
    \item U.S. institutes to support students and faculties from Africa and Latin American to come to U.S. laboratories and universities for research experience programs with short-term visits (three to six months);
    \item For impact assessment, U.S. funding agencies, laboratories and universities support the aforementioned recommendations in a coordinated program managed through a laboratory.
    \item Organizations collectively representing U.S. National Laboratories and Universities in international collaborations should recognize the disparity in economic capabilities of countries in Africa and Latin America compared to the U.S., and in order to best support the development of HEP in these countries, should support and lead initiatives for more equitable contributions (e.g. membership and operations fees for participation in large collaboration, conference fee waivers and travel support to U.S. based meetings, etc).
    \item The Division of Particles and Fields, of American Physical Society, as the organiser for Snowmass, to oversee the implementation of the recommendations between Snowmass 2021 and the next one.
\end{itemize}


\bibliographystyle{elsarticle-num}
\bibliography{myreferences} 

\begin{thebibliography}{10}
\expandafter\ifx\csname url\endcsname\relax
  \def\url#1{\texttt{#1}}\fi
\expandafter\ifx\csname urlprefix\endcsname\relax\def\urlprefix{URL }\fi
\expandafter\ifx\csname href\endcsname\relax
  \def\href#1#2{#2} \def\path#1{#1}\fi

\bibitem{PeaceCorps}
\url{https://history.house.gov/Historical-Highlights/1951-2000/The-establishment-of-the-Peace-Corps}.

\bibitem{USAID}
\url{https://www.usaid.gov/who-we-are/mission-vision-values,
  https://www.usaid.gov/news-information/congressional-testimony/may-26-2021-written-statement-usaid-administrator-samantha-power-senate-fy-2022,
  https://www.usaid.gov/who-we-are}.

\bibitem{aai}
{The African-American Institute}, \url{https://www.aaionline.org/}.

\bibitem{fulbright}
\url{https://us.fulbrightonline.org/}.

\bibitem{BillMelinda}
\url{https://nationalfund.org/investor/bill-melinda-gates-foundation/}.

\bibitem{yale}
\url{https://pier.macmillan.yale.edu/}.

\bibitem{ucla}
\url{https://www.international.ucla.edu/lai/outreach}.

\bibitem{title6}
{The Title VI Act}, \url{https://www.hsdl.org/?view&did=484271}.

\bibitem{UN}
\url{https://iupap.org/2021/12/05/the-international-year-of-basic-sciences-for-sustainable-development-proclaimed-by-the-united-nations-general-assembly-for-2022/}.

\bibitem{CERN}
{CERN}, \url{https://home.cern/}.

\bibitem{IUPAP}
{IUPAP}, \url{https://iupap.org/}.

\bibitem{UNESCO}
{UNESCO}, \url{https://en.unesco.org/}.

\bibitem{UN-indices}
\url{https://www.un.org/development/desa/dpad/least-developed-country-category/ldc-criteria.html}.

\bibitem{WorldBank-indices}
\url{https://worldmapper.org/maps/absolute-poverty-2016/?_sft_product_cat=people,
  https://worldmapper.org/maps/gni-2018/?sf_action=get_data&sf_data=results&_sft_product_cat=economy&sf_paged=2,
  https://worldmapper.org/maps/gdp-2018/?sf_action=get_data&sf_data=results&_sft_product_cat=economy&sf_paged=9}.

\bibitem{Alternate-indices}
\url{https://en.wikipedia.org/wiki/Human_Development_Index,
  https://en.wikipedia.org/wiki/Multidimensional_Poverty_Index,
  https://en.wikipedia.org/wiki/Happy_Planet_Index}.

\bibitem{EconoSectors}
\url{https://en.wikipedia.org/wiki/Three-sector_model}.

\bibitem{cern-map}
\url{https://cds.cern.ch/record/2708601?ln=en}.

\bibitem{APS-taskforce}
{American Physical Society Task Force on Expanding International Engagement},
  \url{https://www.aps.org/programs/international/upload/APS_TaskForceReport_AC.pdf}.

\bibitem{dune}
{The DUNE collaboration map by participating countries},
  \url{https://lbnf-dune.fnal.gov/}.

\bibitem{Allotey1996}
F.~K.~A. Allotey,
  \href{{https://inis.iaea.org/collection/NCLCollectionStore/_Public/28/022/28022232.pdf}}{{Physics
  in Africa}}, {International Atomic Energy Agency} IC/96/166 (1996) 1--19.
\newline\urlprefix\url{{https://inis.iaea.org/collection/NCLCollectionStore/_Public/28/022/28022232.pdf}}

\bibitem{HEP-Africa}
{HEP Activities in Africa},
  \url{https://indico.fnal.gov/event/44870/contributions/195031/attachments/135723/168964/HEPP_Activities_in_Africa_and_Middle_East.pdf}
  (2020).

\bibitem{LHCb}
{Algeria in LHCb},
  \url{https://lhcb.web.cern.ch/lhcb_page/collaboration/organization/list_of_members/members_default.pdf}.

\bibitem{Antares}
{Astronomy with a Neutrino Telescope and Abyss environmental RESearch},
  \url{https://antares.in2p3.fr/}.

\bibitem{KM3Net}
{KM3Net the next generation neutrino telescopes},
  \url{https://www.km3net.org/}.

\bibitem{SA-JINR}
{SA-JINR Collaboration}, \url{https://tlabs.ac.za/iri-g/sa-jinr/}.

\bibitem{iaea}
\url{https://www.iaea.org/}.

\bibitem{AAI-SOE}
{State of Education in Africa},
  \url{https://www.aaionline.org/state-of-education-on-africa-conference/,
  https://www.aaionline.org/state-of-education/}.

\bibitem{ASP}
{The African School of Physics}, \url{https://www.africanschoolofphysics.org/}.

\bibitem{ASESMA}
{African School for Electronic Structure Methods and Applications (ASESMA)},
  \url{https://www.aps.org/publications/apsnews/201612/international.cfm}
  (2016).

\bibitem{ICTP}
{The International Center for Theoretical Physics}, \url{https://www.ictp.it/}.

\bibitem{AIMS}
{The African Institute for Mathematical Sciences},
  \url{https://nexteinstein.org/}.

\bibitem{Cecire2021}
{Enrique Arce-Lareta, K\'et\'evi Assamagan, Emanuela Barzi, Uta Bilow, Kenneth
  Cecire, Sijbrand de Jong, Simone Donati, Steven Goldfarb, Joel Klammer,
  Azwinndini Muronga, Maria Niland}, {The necessities for international HEP
  opportunities for American education}, {arXiv:2203.09336} (2022).
\newblock \href {https://doi.org/https://doi.org/10.48550/arXiv.2203.09336}
  {\path{doi:https://doi.org/10.48550/arXiv.2203.09336}}.

\bibitem{ASP-reports}
{Activity reports of African School of Physics},
  \url{http://africanschoolofphysics.web.cern.ch/2010/asp2010.pdf,
  https://africanschoolofphysics.web.cern.ch/asp2012/asp2012_final.pdf,
  https://www.africanschoolofphysics.org/wp-content/uploads/2014/11/asp2014.pdf,
  https://www.africanschoolofphysics.org/wp-content/uploads/2019/08/ASP2016-FinalReport.pdf,
  https://www.africanschoolofphysics.org/wp-content/uploads/2019/08/ASP2018.pdf}
  (2010-2018).

\bibitem{HEP-LatinAmerica}
{HEP Activities in Latin America},
  \url{https://indico.fnal.gov/event/44870/contributions/195030/attachments/135650/168351/RosenfeldSnowmass2020.pdf}
  (2020).

\bibitem{tandar}
\url{http://www.tandar.cnea.gov.ar} (in Spanish) (2021).

\bibitem{tandar2}
N.~Fazzini, H.~Gonzalez, S.~Tau, A.~Tersigni, Nucl. Inst. Meth. A268~(2) (1988)
  330--333.
\newblock \href {https://doi.org/10.1016/0168-9002(88)90529-3}
  {\path{doi:10.1016/0168-9002(88)90529-3}}.

\bibitem{Oliveira:2019yyx}
{Oliveira, J. R. B.}, {Status and activities of the LAFN (Laborat\'orio Aberto
  de F\'\i{}sica Nuclear)}, J. Phys. Conf. Ser. 1291~(1) (2019) 012003.
\newblock \href {https://doi.org/10.1088/1742-6596/1291/1/012003}
  {\path{doi:10.1088/1742-6596/1291/1/012003}}.

\bibitem{LNLS}
\url{www.lnls.cnpem.br}.

\bibitem{AugerObs}
\url{https://www.auger.org}.

\bibitem{hawc}
\url{https://www.hawc-observatory.org}.

\bibitem{ctao}
\url{https://www.cta-observatory.org}.

\bibitem{swgo}
\url{https://www.swgo.org}.

\bibitem{condor}
\url{https://condorobservatory.ucr.edu}.

\bibitem{Bertou:2012fk}
{Bertou, X.}, {The ANDES underground laboratory}, Eur. Phys. J. Plus 127 (2012)
  104.
\newblock \href {https://doi.org/10.1140/epjp/i2012-12104-1}
  {\path{doi:10.1140/epjp/i2012-12104-1}}.

\bibitem{Dib:2015rma}
{Dib, Claudio O.}, {ANDES: An Underground Laboratory in South America}, Phys.
  Procedia 61 (2015) 534--541.
\newblock \href {https://doi.org/10.1016/j.phpro.2014.12.118}
  {\path{doi:10.1016/j.phpro.2014.12.118}}.

\bibitem{andeslab2}
\url{https://www.andeslab.org}.

\bibitem{LASF4RI2021}
{Latin American Strategy for Research Infrastructures for High Energy,
  Cosmology, Astroparticle Physics LASF4RI for HECAP},
  \url{https://arxiv.org/pdf/2104.06852.pdf} (2021).

\bibitem{HELEN}
\url{https://www.roma1.infn.it/exp/helen/}.

\bibitem{CLASHEP}
{CERN Latin-American School of High Energy Physics},
  \url{https://physicschool.web.cern.ch/clashep/default.html}.

\bibitem{CRCERN2014}
{CERN}, {Proposal for an international cooperation agreement with Costa Rica},
  \url{https://cds.cern.ch/record/1602427/} (2014).

\bibitem{CRCERN2017}
{Costa Rican Legislative Assembly}, {Proposal for an international cooperation
  agreement with Costa Rica},
  \url{http://www.pgrweb.go.cr/scij/Busqueda/Normativa/Normas/nrm_texto_completo.aspx?param1=NRTC&nValor1=1&nValor2=84272&nValor3=108668&strTipM=TC}
  (2017).

\bibitem{MinisterialDeclarationLASF4RI}
{Ministerial Declaration from the Third Ministerial Meeting of Science and
  Technology of Ibero-America},
  \url{https://www.segib.org/wp-content/uploads/Declaracion-III-Reunion-de-Ministros-y-Altas-Autoridades-en-Ciencia-Tecnolog--a-e-Innovacion_ES.pdf}
  (2018).

\bibitem{LA-Strategy}
{Latin American Strategy for Research Infrastructures for High Energy,
  Cosmology, Astroparticle Physics LASF4RI for HECAP},
  \url{https://arxiv.org/pdf/2104.06852.pdf} (2021).

\bibitem{SA-physics}
{Shaping the Future of Physics in South Africa},
  \url{https://www.aps.org/publications/apsnews/200801/internationalnews.cfm}
  (2008).

\bibitem{ASP2021-report}
{K\'et\'evi A. Assamagan, et al.}, {Activity report of the African School of
  Physics, 2019-2021},
  \url{https://arxiv.org/ftp/arxiv/papers/2109/2109.00509.pdf} (2019-2021).

\bibitem{atlas}
{The ATLAS Collaboration}, {ATLAS A 25-year inside story}, {Advanced Series on
  Directions in High Energy Physics} {ISSN 1793-1339; vol 30} (2019).

\bibitem{Egypt-CERN}
{Egypt in CMS and ALICE},
  \url{https://international-relations.web.cern.ch/stakeholder-relations/states/egypt}.

\bibitem{SA-ATLAS}
{South Africa joins ATLAS},
  \url{https://atlas-service-enews.web.cern.ch/2010/news_10/news_SouthAfrica
  joins ATLAS.php}.

\bibitem{An-Najah}
{An-Najah: the First University in Palestine to Join the ATLAS Experiment},
  \url{https://sci.najah.edu/en/news-and-activities/2022/03/najah-first-university-palestine-join-atlas-experiment-cern/}.

\end{thebibliography}

\end{document}